\documentclass[twocolumn,aps,prd]{revtex4-2}
\usepackage{graphicx}
\usepackage{bm}
\usepackage{relsize}
\usepackage{multirow}
\usepackage{xfrac}
\begin{document}
\newcommand*{\bi}{\bibitem}
\newcommand*{\ea}{\textit{et al.}}
\newcommand*{\eg}{\textit{e.g.}}
\newcommand*{\ie}{\textit{i.e.}}
\newcommand*{\plb}[3]{Phys.~Lett.~B \textbf{#1}, #2 (#3)}
\newcommand*{\phrc}[3]{Phys.~Rev.~C~\textbf{#1}, #2 (#3)}
\newcommand*{\phrd}[3]{Phys.~Rev.~D~\textbf{#1}, #2 (#3)}
\newcommand*{\phrl}[3]{Phys.~Rev.~Lett.~\textbf{#1}, #2 (#3)}
\newcommand*{\pr}[3]{Phys.~Rev.~\textbf{#1}, #2 (#3)} 
\newcommand*{\npb}[3]{Nucl.~Phys.~B \textbf{#1}, #2 (#3)} 
\newcommand*{\ptp}[3]{Prog. Theor. Phys. \textbf{#1}, #2 (#3)}
\newcommand*{\prpt}[3]{Phys. Rep. \textbf{#1}, #2 (#3)}
\newcommand*{\ijmpa}[3]{Int. J. Mod. Phys. A \textbf{#1}, #2 (#3)}.
\newcommand*{\epjc}[3]{Eur. Phys. J. C \textbf{#1}, #2 (#3)}. 
\newcommand*{\ra}{\rightarrow}
\newcommand*{\dd}{D\bar D}
\newcommand*{\dpdm}{D^+D^-}
\newcommand*{\dndn}{D^0\bar D^0}
\newcommand*{\dspdsm}{D_s^+D_s^-}
\newcommand*{\kpkm}{K^+K^-}
\newcommand*{\knkn}{K^0\bar K^0}
\newcommand*{\pds}{\psi(2\mathrm S)}
\newcommand*{\rf}[1]{(\ref{#1})}
\newcommand*{\be}{\begin{equation}}
\newcommand*{\ee}{\end{equation}}
\newcommand*{\bea}{\begin{eqnarray}}
\newcommand*{\eea}{\end{eqnarray}}
\newcommand*{\nl}{\nonumber \\}
\newcommand*{\die}{e^+e^-}
\newcommand*{\jj}{\mathrm i}
\newcommand*{\cndf}{\chi^2/\mathrm{DOF}}
\newcommand*{\ndf}{\mathrm{DOF}}
\newcommand*{\minuit}{\texttt{MINUIT}~}
\newcommand*{\e}[1]{{\mathrm e}^{#1}}
\newcommand*{\dek}[1]{\!\times\!10^{#1}}
\newcommand*{\rd}{\mathrm d}
\newcommand{\p}{$p$-value }

\title{The $\psi(2\mathrm S)$ state as a subthreshold pole in 
electron‐positron annihilation into ${D\bar D}$ final states.}
\author{Peter Lichard}
\affiliation{Institute of Physics and Research Centre for 
Computational Physics and Data Processing, Silesian University in Opava,
 746 01 Opava, Czech Republic}

\begin{abstract}
By performing a fit to the merged BES 2008 and BESIII 2024 $e^+e^-\to D\bar D$ 
cross-section data, we find that the $\psi(2\mathrm S)$ subthreshold pole 
influences this process markedly. The statistical significance of the 
$\psi(2\mathrm S)$ as a subthreshold pole is 8$\sigma$. A fit assuming it 
and seven resonances excels with the fit quality characterized by 
$\chi^2/\mathrm{DOF}=240.4/284$, which means a $p$-value of 97\%. The 
interference of resonances with the continuum provided by the subthreshold 
pole changes their shapes, positions, and widths.
\end{abstract}
\maketitle

\section{Introduction}
\label{introduction}

Subthreshold poles are very well-known in hadronic reactions. 
A typical example is the nucleon pole in the amplitude of the pion-nucleon 
scattering. Even if it is not accessible in an experiment, it strongly 
influences the cross section. Owing to that, its existence can be proven 
by analyzing the forward scattering amplitude obtained from the phase-shift 
analyses, and its residue proportional to the $\pi$NN coupling constant 
can be determined \cite{pin1,pin2}. 

To our knowledge, nobody has systematically investigated the analytic 
properties of the amplitudes of the $\die$ annihilation to various hadronic 
systems. Here, we will assume that they are similar to those of hadronic
amplitudes, discussed, \eg, in \cite{analprop1,analprop2,analprop3}. 
From general principles of causality, locality of interactions, and unitarity, 
the amplitude is a real analytic function in the complex $s$-plane with the cut
along the positive real axis running from the process threshold to infinity
(called the physical cut). 
The stable states, i.e., those which do not decay into the considered final 
state, are represented by subthreshold poles (SP) lying on the real axis below 
the threshold. 
Resonances are represented by pairs of complex-conjugate poles on the higher 
Riemann sheets, accessible through the physical cut. 
Their contributions to the amplitude on the upper branch of the 
physical cut are usually parametrized by some form of the Breit-Wigner 
formula.

Vector-meson dominance (VMD) \cite{vmd1,vmd2,vmd3,vmd4} is a useful 
phenomenological concept 
based on the assumption that the interaction of the hadronic system with the 
electromagnetic field is mediated by truly neutral vector meson resonances 
of negative C-parity. It is frequently used to interpret the $\die$ 
annihilation data into hadrons. The intermediate vector mesons with masses 
above the considered reaction's threshold produce salient features of 
excitation curves (peaks, bumps, dips, steep slopes). Those with masses
below the threshold should appear as SPs.

In this paper, we show that achieving a satisfactory fit to the 
$e^+e^-\to D\bar D$ data requires an SP pole in the fitting formula, 
indicating that an SP is present in the process amplitude. 

The experimental data we explore are presented in the next section. Our model 
is described in Section \ref{model}. An SP crawling in and its identification 
as $\pds$ are detailed in Section \ref{SPappears}. The ultimate fit with the 
$\pds$ as a subthreshold pole and seven resonances is reported in Section 
\ref{results}. The role of the two factors contributing to the high-goodness 
fit is analyzed in Section \ref{discussion}. We recapitulate in Section 
\ref{conclusions}.

\section{The explored data}
The experimentalists working at the spectrometer BESIII, situated at 
the electron-positron collider BEPCII, based at the IHEP laboratory
in Beijing, China, have recently published \cite{besiii2024} precise data 
on $\die$ annihilation to the $\dd$ final states. In the Supplemental
Material \cite{suppl}, they provide the Born cross sections of the
$\die\ra\dndn$ and $\die\ra\dpdm$ processes at 150 
center-of-mass system (CMS) energies between 3.80 and 4.95 GeV. 
Additionally, correction factors are tabulated, enabling the conversion 
of Born cross sections to dressed cross sections.
The dressed cross sections were the subject of the fit presented in the 
Supplemental Material \cite{suppl} and resulted in the \p of 0.2\%. 
We will follow the experimentalists and use the dressed cross sections also 
in this work.

Unfortunately, the BESIII data \cite{besiii2024} do not cover the 
region between the threshold and 3.8~GeV, which contains the $\psi(3770)$ 
resonance. The region above the threshold is expected to be most influenced
by a subthreshold pole. For that reason, we searched for additional
data to cover this region as well. 

In 2018, the BESIII Collaboration published \cite{besiii2018}
the values of the $\dndn$ and $\dpdm$ cross sections only at the 
$\psi(3770)$ peak, even if some work within the Collaboration had been done on 
the whole resonance region \cite{julin}. Finally, we found guidance in Ref.
\cite{ShamovTodyshev}, where a careful analysis of this region was done. 
We have chosen data from the BES Collaboration \cite{bes2008}, 
taken at the same IHEP laboratory, but at the BEPC collider and with the 
conventional magnetic detector, BES-2. Their Table 2 presents the observed
(``dressed'') cross section of both processes of interest at 14 energy points. 

We merged the BES and BESIII data and will determine the free parameters of 
our model by fitting the formula \rf{sigma} below to the combined dataset
containing 328 points.


\section{Model}
\label{model}

\begin{figure}
\centering
\includegraphics[width=0.39\textwidth,height=0.13\textwidth]{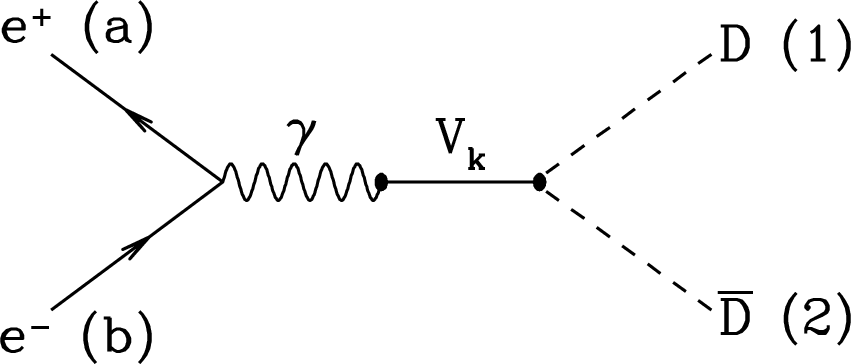}
\caption{\label{fig:ee2ddbar}Feynman diagram defining our model.}
\end{figure}

For the description of the electron-positron annihilation into a $\dd$ pair
($D$ represents $D^+$ or $D^0$), we use a VMD model 
based on the Feynman diagram depicted in Fig.~\ref{fig:ee2ddbar} and the 
interaction Lagrangian
\be
\label{lagr}
{\cal
L}_{V\!\phi}(x)=\jj{g_{V_k\!\phi}}V_\mu(x)\left\{[\partial^\mu\phi^\dagger(x)]
\phi(x)-\phi^\dagger(x)\partial^\mu\phi(x)\right\},
\ee
where $V^\mu(x)$ denotes the (hermitian) vector field, the quantum of which
is the $V_k$ meson, and $\phi(x)$ the pseudoscalar field with quanta $D$ and 
$\bar D$. The $\gamma V_k$ junction is parametrized as 
$eM_k^2/g_k$ in analogy with the $\gamma\rho^0$ junction 
$eM^2_{\rho^0}/g_\rho$. The formula for the cross section of the $\die$ 
annihilation into a $\dd$ pair based on the VMD model 
with $n$ resonances is derived in the Appendix. It sounds

\be
\label{sigma}
\sigma(s)=\frac{\pi\alpha^2}{3}\frac{\beta^3}{s}
\left|\sum_{k=1}^n
\frac{\sqrt{Q_k}\,e^{\jj\delta_k}}{s-M_k^2+\jj M_k\Gamma_k}
\right|^2\!.
\ee
In this formula, $M_k$ and $\Gamma_k$ are the mass and width of the $k$th 
resonance, respectively, $Q_k=R_k^2$, where $R_k=M_k^2g_{V_k\phi}/g_{V_k}$, 
$\delta_k$ is an additional phase shift, and
\be
\label{beta}
\beta=\sqrt{1-\frac{4m_D^2}{s}}
\ee 
is the speed of the $D$ meson in CMS.

Formula \rf{sigma} is strictly valid for the Born cross section. However,  
the Tables 1 and 2 in \cite{suppl} show that the virtual polarization factor 
$|1-\Pi|^{-2}$ \cite{actis} is practically constant in the energy range we 
deal with (it varies between 1.04 and 1.06). We will absorb its mean value into 
$Q_k$ and use the same formula for the dressed cross section. 

We will treat $M_k$s, 
$\Gamma_k$s, $Q_k$s, and $\delta_k$s as free parameters  (except $\delta_1$, 
which is kept at 0). We will determine their mean values and dispersions by 
fitting the formula \rf{sigma} to experimental cross sections using the 
standard $\chi^2$ criterion \cite{footminuit}.

In the seminal paper \cite{multiple}, the mathematical structure of the 
relativistic Breit-Wigner function was analyzed. The authors found that there 
are $2^{n-1}$ fitting solutions with equal quality for $n$ resonances with 
masses $M_k$ and widths $\Gamma_k$, $k=1,\ldots,n$, differing in multiplicative 
parameters (in this paper, $Q_k$) and phase shifts $\delta_k$. Because our 
argument about the existence of SP does not rely on specific values of those 
parameters, we will avoid investigating the phenomenon of multiple solutions.

\section{A subthreshold pole appears}
\label{SPappears}
The BESIII Collaboration \cite{besiii2024} assumed in their fit that all 
parameters of the $\die\ra\dpdm$ and $\die\ra\dndn$ processes are the same.
Here, we will consider the differences between the two processes. 
Their origins are physical (different masses of $D^+$ and 
$D^0$ and possibly different couplings to the intermediate vector
mesons), as well as experimental (different acceptances rooted in different 
methods of identifying particles in final states).
We will split each of the $R_k$ and $\delta_k$ parameters into two. Those 
pertinent to the $\dndn$ final states will be marked by subscript $a$; those 
to the $\dpdm$ ones by $b$. We will assume that the resonances' parameters 
$M_k$ and $\Gamma_k$ are the same for both processes.

To help the minimization code cope with many parameters, we started the
fitting with two resonances. We gradually added further, using the already
found parameters as the starting ones for the next run. 

In this way, we reached the six-resonance fit with $\cndf=370.6/294$ and a 
\p of 0.16\%. The details of the fit are shown in Table \ref{tab:6reso_q}.
\begin{table*}[h]
\centering
\begin{tabular}{lcccccc}
\hline 
\multicolumn{6}{c}{$\chi^2/$NDF=370.6/294~~~~~~~~~~~~~$p$-value$=0.2$\%}\\
\hline
$k$ & 1&2 & 3 & 4 & 5 & 6\\ 
\hline
$M$ (MeV)&$3761.1\pm2.1$&$3778.9\pm1.0$&$4028.7\pm1.1$&$4071.4\pm3.5$&
$4210.2\pm2.5$&$4420.3\pm4.3$\\
$\Gamma$ (MeV)&$123.4\pm4.4$&$35.7\pm1.0$&$66.3\pm2.9$&$452\pm16$&
$34.3\pm3.8$&$102.0\pm6.6$\\
$Q_a$ (GeV$^4$)&$134.9\pm4.8$&$24.1\pm1.7$&$0.570(78)$&$37.6\pm1.6$&
$2.65(91)\dek{-3}$&0.061(12)\\
$Q_b$ (GeV$^4$)&$131.2\pm4.0$&$22.2\pm1.5$&0.541(71)&$39.8\pm1.6$&
$3.08(90)\dek{-3}$&0.0450(90)\\
$\delta_a$ (rad)&-0.310(22)&2.507(45)&0(f)&-2.629(23)&-2.42(23)&-1.17(24)\\
$\delta_b$ (rad)&-0.298(19)&2.526(40)&0(f)&-2.613(22)&-2.37(23)&-1.29(21)\\
\hline
\end{tabular}
\caption{\label{tab:6reso_q}Parameters of the joint fit to the merged
$\dndn$ (subscript $a$) and $\dpdm$ (subscript $b$) data using Eq. \rf{sigma} 
with the six resonances. The masses $M$ and widths $\Gamma$ are assumed to be 
the same in both processes.}
\end{table*}

To improve the fit, we added a new resonance with unconstrained
parameters. The quality of the fit increased to $\cndf=289.2/288$,
\p= 47\%. The statistical significance of the new resonance was 7.9$\sigma$,
estimated by utilizing the changes in likelihood values $\delta(-2\ln
L)=81.4$ and in the number of degrees of freedom $\delta(\ndf)=6$.

But the parameters of the new ``resonance'' are surprising: $M=(3691\pm56)$~MeV
and $\Gamma=(0\pm160)$~MeV. They describe not a usual Breit-Wigner resonance, 
but an SP lying on the real axis of the complex $s$-plane
below the reaction threshold of 3729.68(10)~MeV [based on the D$^0$ mass
from \cite{pdg2026}]. We tried to bring the minimizing code to sense by 
setting the minimal resonance mass slightly above the threshold. As a result, 
the $\chi^2$ minimization gave one resonance with the mass identical 
to the chosen limit and a $\chi^2$ value higher than that with the SP. So it 
was impossible to finish the work with only the resonances, and we had to 
accept the presence of an SP.

When trying to identify the SP we have just discovered with an existing
state, we should not be fooled by its zero width. What looks like a stable 
state from the point of view of the $\dd$ final states may decay in other
processes. There are two obvious candidates with the $J^{PC}=1^{--}$ quantum 
numbers and the quark composition ensuring the strong coupling to the 
$D\bar D$ system: $J/\psi(1\mathrm S)$ with mass about 3097 MeV and $\pds$ with
3686.097(11) MeV \cite{pdg2026}. The mass of the SP we have found is by 
(5$\pm$57)~MeV higher than that of the $\pds$ and compatible with it. This 
fact leads us to choose the $\pds$ for further processing.

\section{Results}
\label{results}
We reperform a fit with six resonances and an SP with the mass fixed at that
of the $\pds$ and vanishing width. The goodness of the fit is characterized
by $\cndf=289.5/290$ and \p= 50\%. Comparison with the fit with six
resonances only implies the $\pds$ statistical significance as an SP of 
8.2$\sigma$ (based on $\delta(-2\ln L)=81.1$ and $\delta(\ndf)=4$).

To further improve the fit quality, we added another resonance, bringing
the total number of resonances accompanying the $\pds$ SP to seven. This move 
has resulted in $\cndf=240.4/284$ and \p= 97\%. The contributions of the $\dpdm$
and $\dndn$ channels to the $\chi^2$ are 119.5 and 120.9, respectively. 
Statistical significance against the previous solution, calculated from 
$\delta(-2\ln L)=49.1$ and $\delta(\ndf)=6$, is 5.7$\sigma$.

We show the detailed results in Table \ref{tab:duo_parms} \cite{attempts}.
In addition
to the parameters of the basic formula \rf{sigma}, we also present the local
significances of the SPs in the $\dpdm$ and $\dndn$ amplitudes and those
of resonances. We estimated them from the $Q_k$ coefficient's central value
and dispersion as $S_k=Q_k/\sigma_{Q_k}$.
\begin{table*}[b]
\begin{tabular}{lcccccccc}
\hline 
\multicolumn{8}{c}{$\cndf=240.4/284$~~~~~~~~~~~~~~~~~\p= 97\%}\\
\hline
$k$ & $1\,\equiv \pds$ & $2\,\equiv\psi(3770)$ & 3 & 4 & 5 & 6 & 7 & 8 \\
\hline
$M$ (MeV) &~3686.097(f) & 3779.13(76) & 3897$\pm$13 & 4026.6$\pm$1.9 &
 4101.7$\pm$6.3 & ~4207.8$\pm$3.3 & 4408.6$\pm$6.1 & 4572.8$\pm$6.8 \\
$\Gamma$ (MeV)&  0(f) & 27.1$\pm$1.6 & 149$\pm$16 & 29.4$\pm$3.1 &
94$\pm$20 & 41.7$\pm$5.4  & 110$\pm$12 & 46$\pm$14\\
$Q_a$ (GeV$^4$) & $52.2\pm6.6$ & 6.42(84) & 2.39(96) &0.134(35)&
0.16(11) & 0.0045(18)& 0.035(11) & 1.15(86)$\dek{-3}$  \\
$Q_b$ (GeV$^4$) & $53.7\pm6.3$ &5.81(76) & 2.06(79) & 0.131(34)&
0.139(94)& 0.0048(17) & 0.0261(81) & 0.96(64)$\dek{-3}$ \\
$\delta_a$ (rad)& 0(f)& $-2.757(93)$ & 2.44(25) & $-0.80(18)$&
1.95(23) &2.30(25) &1.21(13) & -0.88(30) \\
$\delta_b$ (rad)& 0(f)&$-2.729(88)$ & 2.44(27) & $-0.87(15)$ &
 1.98(21) & 2.20(25) & 0.98(13) & -1.44(31)\\
$S_a$ & 7.9$\sigma$ &7.6$\sigma$ &2.5$\sigma$ &3.8$\sigma$ &1.5$\sigma$ 
&2.5$\sigma$ &3.2$\sigma$ & 1.3$\sigma$ \\
$S_b$ & 8.5$\sigma$ &7.6$\sigma$ &2.6$\sigma$ &3.9$\sigma$ &1.5$\sigma$ 
&2.8$\sigma$ & 3.2$\sigma$ & 1.5$\sigma$ \\
\hline
\end{tabular}
\caption{\label{tab:duo_parms}Parameters of the joint fit to the merged
$\dndn$ (subscript $a$) and $\dpdm$ (subscript $b$) data 
using Eq. \rf{sigma} with the $\pds$ SP and seven 
resonances. The masses $M$ and widths $\Gamma$ are assumed to be 
the same in both processes. Also, the local significances $S_a$ and $S_b$
are shown.}
\end{table*}

\begin{figure}[h]
\includegraphics[width=0.483\textwidth,height=0.34\textwidth]
{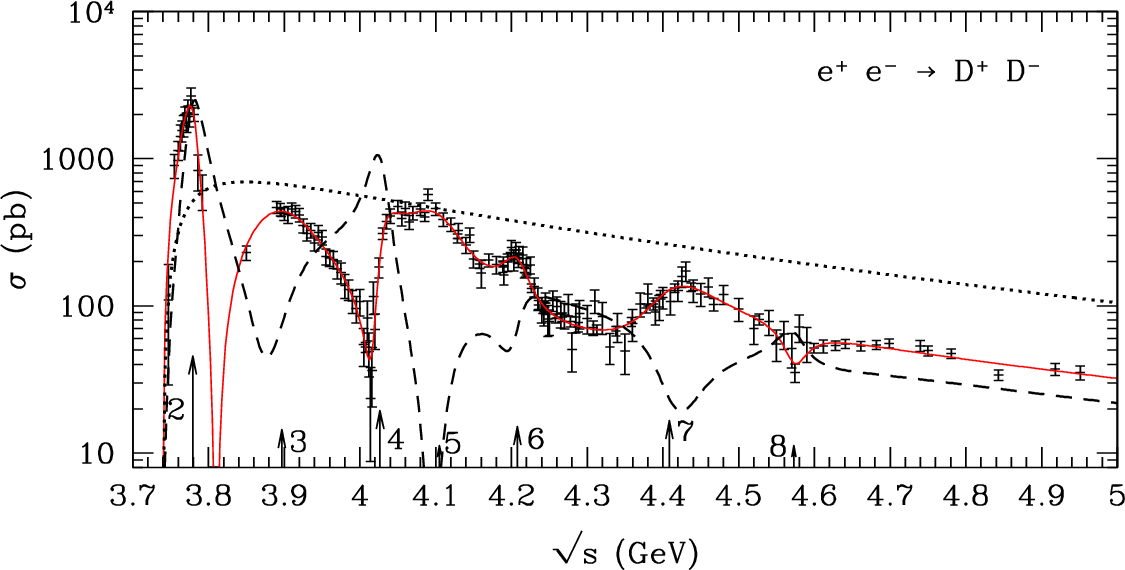}
\caption{\label{fig:DpDm}Full line (red online): The $\die\ra\dpdm$ cross 
section as it follows from the joint fit to both processes, assuming the $\pds$ 
SP with seven resonances, marked by arrows with their ordinal numbers from 
Table \ref{tab:duo_parms}. The lengths of the arrows characterize local
significances. The lines with error bars represent the merged BES and BESIII 
data. The dotted/dashed curve depicts the cross section if only the 
SP/resonance part is considered.}
\end{figure}

\begin{figure}[h]
\includegraphics[width=0.483\textwidth,height=0.34\textwidth]
{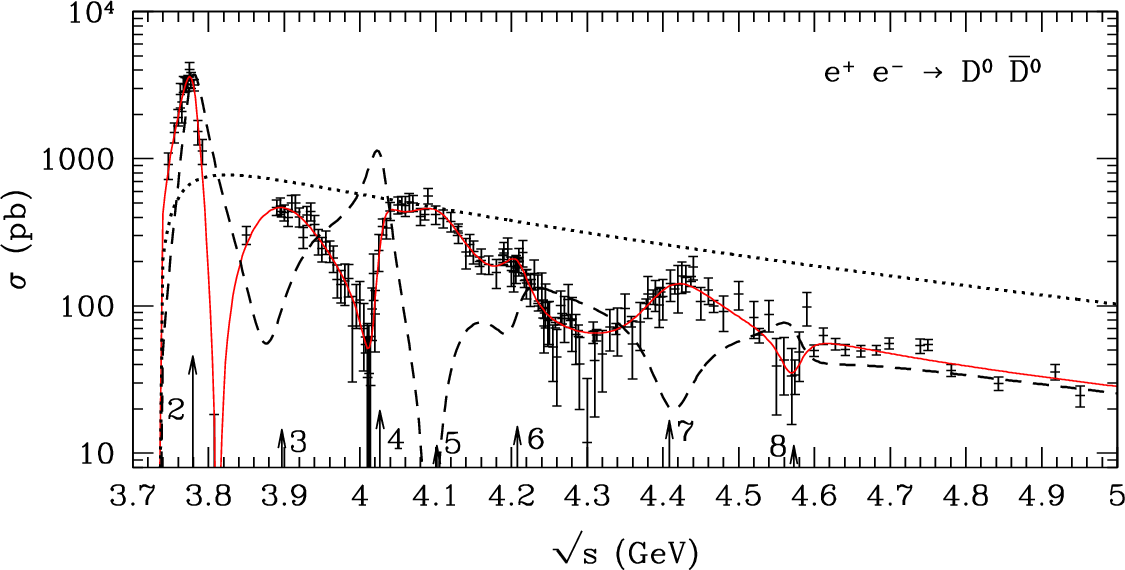}
\caption{\label{fig:D0aD0}The same as Fig. \ref{fig:DpDm}, but for the
$\die\ra\dndn$ process.} 
\end{figure}

The $\die\ra\dpdm$ and $\die\ra\dndn$ cross sections calculated from the 
parameters shown in Table~\ref{tab:duo_parms} are compared to data in 
Fig.~\ref{fig:DpDm} and Fig.~\ref{fig:D0aD0}, respectively. It is necessary
to say that the important experimental points at $\sqrt s=3.80765$~GeV,
which greatly influence the excitation function behavior, are not visible in 
the figures. Their values are \cite{besiii2024}: $\sigma=(0.11\pm7.95)$~pb in 
Fig.~\ref{fig:DpDm} and $\sigma=(4.03\pm14.2)$~pb in Fig.~\ref{fig:D0aD0}.
 

In both Fig. \ref{fig:DpDm} and Fig. \ref{fig:D0aD0}, we show two additional 
curves. The first of them (dotted) represents the cross section
calculated only from the SP part of the amplitude ($k=1$ in Table 
\ref{tab:duo_parms}), ignoring all resonances ($k>1$).
Another (dashed) shows the cross section calculated only from the resonance 
part of the amplitude, \ie, setting $Q_{a,1}=Q_{b,1}=0$ and taking all 
other parameters from Table \ref{tab:duo_parms}.

A striking feature appearing in both processes, see Fig. \ref{fig:DpDm} and 
Fig. \ref{fig:D0aD0}, is the inversion of
all resonances, except the strongest, $\psi(3770)$. The peaks/dips of the 
cross-section curve replace the dips/peaks in the resonance curve.
It is the consequence of the interference between the resonance part of the 
amplitude and the SP pole corresponding to the $\pds$ state.

This phenomenon explains the failure of the BESIII collaboration's fit 
\cite{suppl} assuming seven resonances fixed at the PDG values and not 
considering an SP. In reality, due to the interference with the SP, those 
resonances appeared differently and at different places than expected 
in the fit.

\begin{figure}[h]
\includegraphics[width=0.483\textwidth,height=0.31\textwidth]{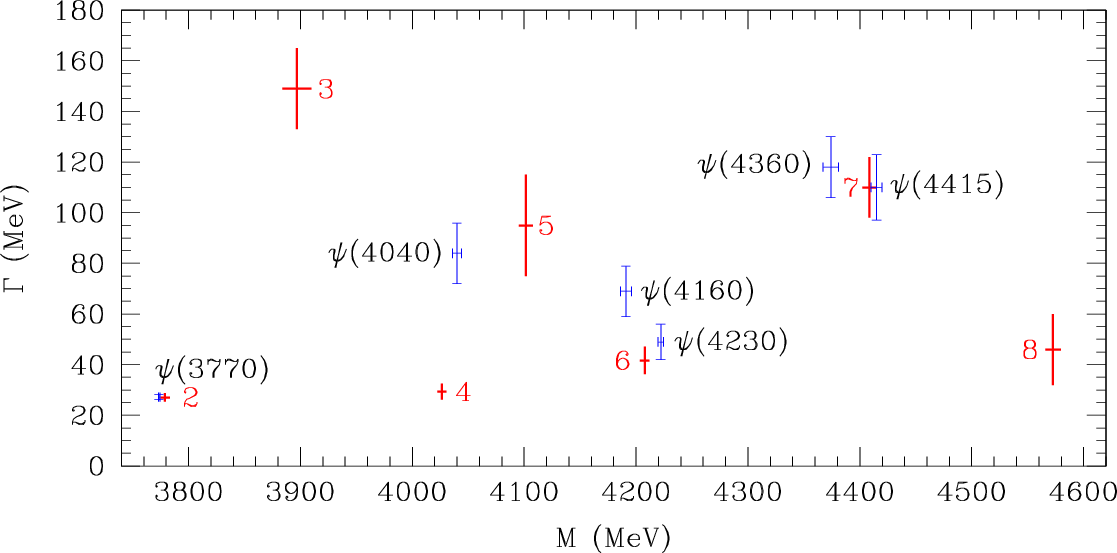}
\caption{\label{fig:mg_psi2S_7R}Comparison of the seven resonances found here
(marked with simple numerals, red online) with those in the PDG 2026 tables, 
distinguished by short lines at endpoints (blue online).}
\end{figure}

It may be instructive to compare the results on resonances that we obtained 
by fitting the very precise BESIII data \cite{besiii2024} supplemented
in the $\psi(3770)$ region by the BES data \cite{bes2008}. To facilitate the
discussion, in Fig. \ref{fig:mg_psi2S_7R} we also provide the $M-\Gamma$ 
plot containing the PDG 2026 \cite{pdg2026} $c\bar c$ resonances together 
with those resulting from
our analysis. The latter are supplied with their ordinal $k$-numbers from Table
\ref{tab:duo_parms}. The same are also used by arrows depicting the
resonance positions in Figs. \ref{fig:DpDm} and \ref{fig:D0aD0}.

The mass 3779.13(76)~MeV and width (27.1$\pm$1.6)~MeV of resonance \#2
agree well with the ``Our Average'' $\psi(3770)$ PDG 2026 values 
(3778.1$\pm$0.7)~MeV and (27.5$\pm$0.9)~MeV, respectively.

The resonance \#3 with $M_3=(3897\pm13)$~MeV and $\Gamma_3=(149\pm16)$~MeV
corresponds to the $G(3900)$ resonance, the nature of which is still debated. 
Our values are compatible with $M=(3898.4\pm0.9)$~MeV and 
$\Gamma=(127.5\pm6.7)$~MeV found in a recent global coupled-channel analysis 
\cite{satoshi}. No mention of $G(3900)$ appears in the PDG 2026 Tables.

The parameters of the $\psi(4040)$ resonance, shown in the PDG Tables
\cite{pdg2026}, originate from one experiment performed a long time ago 
\cite{bes2008a}. Therefore, it may not be too presumptuous to assume that our 
resonance \#4 is a contemporary incarnation of the $\psi(4040)$, based on
the very precise BESIII $\dd$ data \cite{besiii2024}. Moreover, our values 
$M_4=(4026.6\pm1.9)$~MeV, $\Gamma_4=(29.4\pm3.1)$~MeV are compatible with those 
determined by fitting \cite{dspdsm} the very precise $D_s^+D_s^-$ data 
\cite{besiii2024b}, namely $M_4=(4026.7\pm1.7)$~MeV,
$\Gamma_4=(36.4\pm3.3)$~MeV.

The resonance with parameters $M_5=(4101.7\pm6.3)$~MeV, $\Gamma_5=(94\pm20)$~MeV
has both statistical significances of $1.5\sigma$, which qualifies it more
as a fluctuation, not a real physical state.

The resonances $\psi(4160)$ and $\psi(4230)$ have not appeared from our
fit. Instead of them, a resonance with parameters $M_6=(4207.8\pm3.3)$~MeV, 
$\Gamma_6=(41.7\pm5.4)$~MeV has. This is the same situation as with the
precise $\die\ra\dspdsm$ data \cite{besiii2024b} analyzed in Ref. \cite{dspdsm}.
The ``extra'' resonance there has the parameters $M=(4212.9\pm4.7)$~MeV, 
$\Gamma=(48.4\pm8.7)$~MeV, in full conformity with our resonance \#6.
If we, referring to the ``$\dd$ not seen'' in PDG 2026, ignore the
$\psi(4230)$, a problem remains why our \#6 differs so much from the
$\psi(4160)$'s $M=(4191\pm5)$~MeV and $\Gamma=(69\pm10)$~MeV \cite{pdg2026}.

The PDG's $\psi(4360)$ did not show up in our fit. The reason may be its weak
affinity to the $\dd$ channel. We refer to a nonobservance of its $\dd$ decay,
citing the PDG 2026.

Our resonance \#7 perfectly agrees with the PDG's $\psi(4415)$. Just the mass
is a little smaller, $(4408.6\pm6.1)$~MeV against $(4415\pm5)$~MeV.

Resonance \#8 with $M_8=(4572.8\pm6.8)$~MeV, $\Gamma_8=(46\pm13)$~MeV manifests 
itself as a dip in both excitation curves. However, its local significance is 
minor: 1.3$\sigma$ in the $\dpdm$ amplitude and 1.5$\sigma$ in the $\dpdm$ one.
It may be just a fluctuation, without any physical meaning.

\section{Discussion}
\label{discussion}
\begin{figure}
\centering
\includegraphics[width=0.49\textwidth,height=0.35\textwidth]{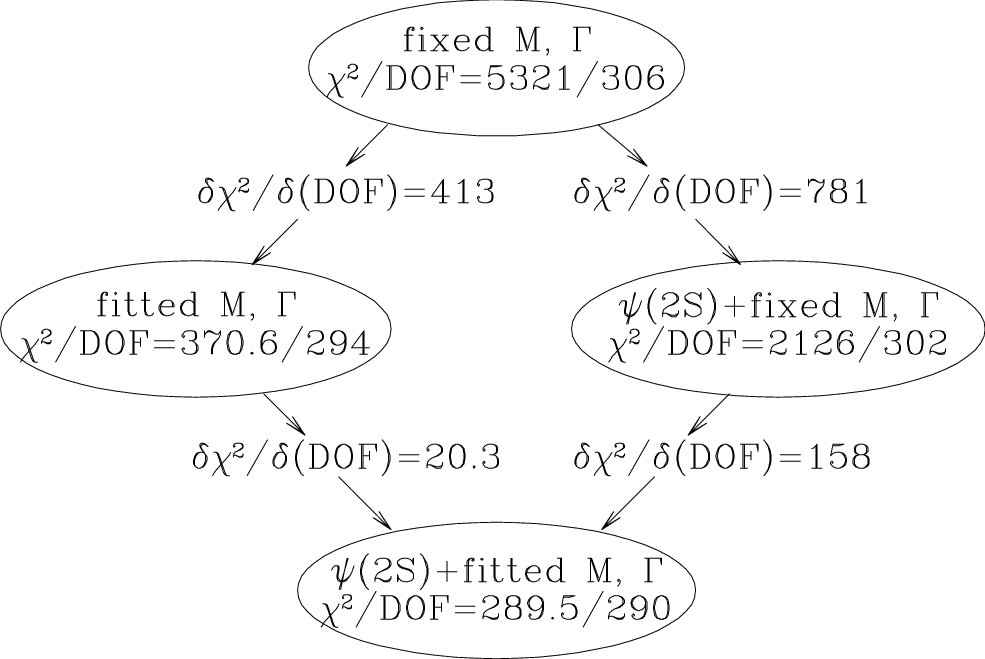}
\caption{\label{fig:psi2S_6R}
Two paths of fitting starting with six
charmonia from the PDG Tables \cite{pdg2026}.}
\end{figure}

There are two factors contributing to our model's agreement with the data.
One of them is fitting the resonances' parameters to the data we explore.
Another is allowing the subthreshold pole with the mass of the $\psi(2S)$
to sneak into the game. To investigate the role of the two factors, we 
conducted a dedicated study, schematically shown in Fig. \ref{fig:psi2S_6R}.

We started fitting with six charmonia, namely $\psi(3770)$, $\psi(4040)$, 
$\psi(4160)$, $\psi(4230)$, $\psi(4360)$, and $\psi(4415)$,  with the masses 
and widths fixed at the PDG values \cite{pdg2026}. Varying other 22 parameters 
($Q$s and $\delta$s), we got the $\chi^2=5321$. 

Then, following the left-hand branch of the scheme, we allow the masses and 
widths of the resonances to vary, which means twelve more parameters. The 
$\chi^2$ has dropped to 370.6. It is impossible to appraise this step in terms 
of statistical significance, so we characterize it by the drop in the $\chi^2$ 
per added parameter. The solution we reached is the same as the six-resonance 
solution obtained in Sec. \ref{SPappears} and summarized in Table 
\ref{tab:6reso_q}. It is the best solution before an SP has intruded, so it is 
worth analyzing it in more detail:
\begin{figure}[h]
\includegraphics[width=0.483\textwidth,height=0.31\textwidth]{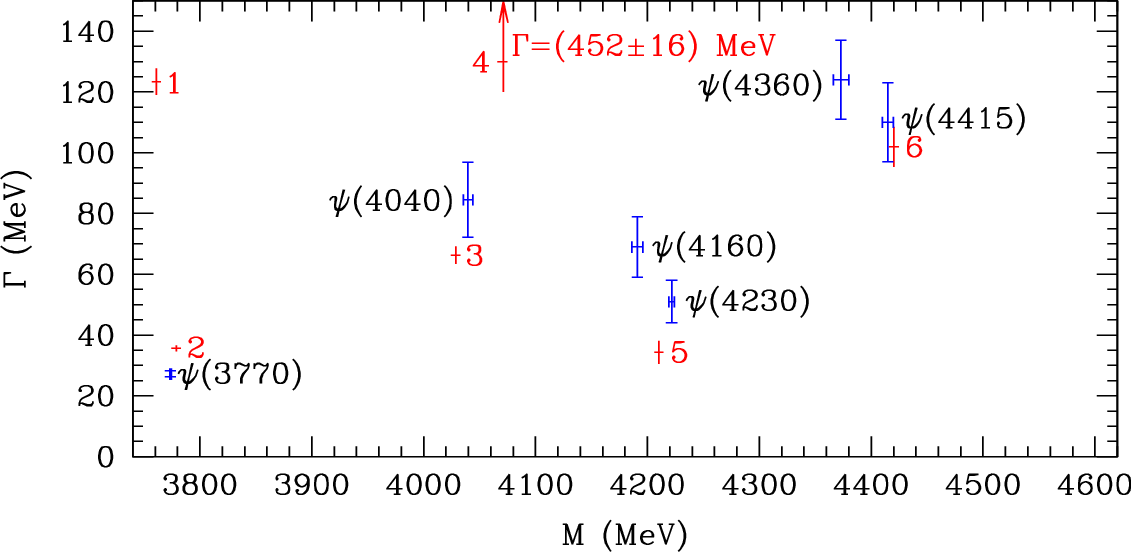}
\caption{\label{fig:mg_6R}Comparison of the six fitted resonances
marked with numerals denoting the ordinal numbers in Table \ref{tab:6reso_q}
(red online) with those in the PDG 2026 tables, 
distinguished by short lines at endpoints (blue online).}
\end{figure}

We compare the resonances found there with established charmonia in Fig. 
\ref{fig:mg_6R}. Besides the $\psi(4415)$, there is no
coincidence. Even the 
width of the salient $\psi(3770)$ strongly ($6\,\sigma$) disagrees with the PDG 
value. The source of the failure lies in fitting the experimental data, which 
are formed by the strong subthreshold pole, resonances, and their interference,
with a fitting formula that contains only resonances. The latter, thus, instead 
of finding the resonances, localized the salient features of the experimental 
excitation curves. The good results for $\psi(4415)$ are achieved 
because it is far from other resonances and the SP contribution varies slowly 
there. Its flip-over is therefore not disturbed. The position of the maximum 
is the same as that of the minimum. 

Continuing along the left-hand branch, we add the subthreshold pole with the 
$\psi(2S)$ mass and vanishing width (four free parameters) and repeat the 
resonance fitting, getting a solution with $\cndf=289.5/290$ and \p of 50\%. 
Here, we do not discuss it in detail, because in Section \ref{results} we 
further improved it by adding the seventh resonance, reaching 
$\cndf=240.4/284$ and \p of 97\%

Following the right-hand branch of the scheme, we included the $\pds$ as an 
SP (four free parameters) into the fitting formula, keeping the resonances' 
masses and widths fixed. The $\chi^2$ dropped from 5321 to a still-very-high 
2126, together with DOF=302, implying a vanishing \p. It is clear that mere 
SP alone is insufficient to achieve a good fit. 

Subsequent fitting of the masses and widths of the resonances yields the same 
final solution as the left-hand branch.

To summarize, we have learned that: (i) Without SP, it is possible to 
achieve a good fit with fitted resonances, perhaps after considering a 
possible line-shape improvement and adding another resonance, but with false 
resonance parameters as a result; (ii) It is impossible to get a good fit 
with an SP and the current PDG resonance masses and widths; (iii) To get a 
reliable and good fit, one has to consider both SP and resonances with all 
parameters fitted.

\section{Summary and conclusions}
\label{conclusions}
The inclusion of an SP significantly improves the quality of a
six-resonance fit to the combined BES \cite{bes2008} and BESIII
\cite{besiii2024} data, expressed by the $p$-value rising from 
0.16\% to 47\% and the SP statistical significance of 7.9$\sigma$. 
The obtained SP position is very close to the $\pds$ resonance mass, which has 
led us to hypothesize that the $\pds$ state, a resonance in other processes, 
behaves in the $\die\ra\dd$ process like a subthreshold pole with a 
statistical significance of 8.2$\sigma$.

Finally, we have supplemented the $\pds$ SP with seven resonances and
performed a fit. It resulted into $\cndf=240.4/284$ and \p= 97\%. 

Surprisingly strong influence of the narrow resonance $\pds$, acting here 
as a subthreshold pole, seen in Figs. \ref{fig:DpDm} and \ref{fig:D0aD0}, 
is caused by its position just below the open-charm threshold and by the 
$D\bar D$ states are the very first ones above it.

All resonances, except for very strong $\psi(3770)$, flip over as a result 
of their interference with the SP (see Figs. \ref{fig:DpDm} and
\ref{fig:D0aD0}). The shape (width) of the $\psi(3770)$ is influenced by 
interference with the rapidly varying SP contribution. Those effects make
the fitting that supposes only resonances untrustworthy.

{\bf A warning:} The widely used version of the Breit-Wigner formula, 
which uses as a variation parameter the product 
${\cal B}(V_k\to D\bar D)\times\Gamma(V_k\ra\die)$, is not able to detect 
an SP because of a negative number appearing under the square root.

\section*{Acknowledgements}
This research did not receive any specific grant from funding agencies in 
the public, commercial, or not-for-profit sectors.

The author thanks Josef Jur\'{a}\v{n} for enlightening him on some statistics
aspects and for valuable remarks on the text, and Veronika Gintnerov\'{a} 
for checking the transfer of information from the computer outputs to the 
manuscript.

The data analyzed in this article are openly available \cite{besiii2024,
bes2008}.

\appendix
\section{Derivation of formula \rf{sigma}}
The Lagrangian \rf{lagr} implies for the $V_k\dd$ vertex in Fig.
\ref{fig:ee2ddbar} the expression $g_{V_k\phi}(p_1-p_2)^\rho$. Together with
the $eM_k^2/{g_k}$ for the $\gamma V_k$ junction and standard Feynman rules,
it leads to the following amplitude of the $\die\ra\dd$ process 
($p=p_a+p_b$, $s=p^2$, electron mass neglected):
\bea
\jj{\cal M}_{fi}&=&\jj e\bar{v}(p_a,s_a)\gamma_\mu u(p_b,s_b)\,
\frac{-\jj g^{\mu\nu}}{s}\,\frac{eM_k^2}{g_k}\nl
&\times&\jj\frac{-g_{\nu\rho}+\frac{p_\nu p_\rho}{M_k^2}}{s-M_k^2+\jj M_k
\Gamma_k}\,g_{V_k\phi}(p_1-p_2)^\rho \ .\nonumber
\eea 
After some editing and introducing the notation 
\be
\label{Rk}
R_k=M_k^2g_{V_k\phi}/g_k, 
\ee
the 
amplitude squared takes the form
\bea
|{\cal M}_{fi}|^2&=&\frac{e^4}{s^2}\left|\frac{R_k}{s-M_k^2+\jj M_k
\Gamma_k}\right|^2\!(p_1-p_2)^\mu(p_1-p_2)^\nu\nl
&\times&{\bar v}(p_a,s_a)\gamma_\mu u(p_b,s_b){\bar u}(p_b,s_b)\gamma_\nu
v(p_a,s_a)\ .  \nonumber
\eea
Averaging over initial spin states leads to [$t=(p_1-p_a)^2$]
\bea
\overline{|{\cal M}_{fi}|^2}&=&\frac{2e^4}{s^2}\left|\frac{R_k}
{s-M_k^2+\jj M_k\Gamma_k}\right|^2\nl
&\times&(-st-t^2+2m_D^2t-m_D^4)\ .\nonumber
\eea
Inserting this into the differential cross section formula 
\[
\frac{\rd\sigma}{\rd t}=\frac{1}{64\pi s}\frac{1}{|\vec p_a|^2}
\overline{|{\cal M}_{fi}|^2}
\]
and integrating over $t$ from $t_1=m_D^2-s(1+\beta)/2$
to $t_2=m_D^2-s(1-\beta)/2$, where $v_D$ is the
$D$ meson speed, and putting $R_k=\sqrt{Q_k}$, we finally get
\be
\label{sigma1V}
\sigma(s)=\frac{\pi\alpha^2}{3}\frac{\beta^3}{s}
\left|
\frac{\sqrt{Q_k}}{s-M_k^2+\jj M_k\Gamma_k}
\right|^2\!.
\ee

Generalizing Eq. \rf{sigma1V} to the case of $n$ interferring resonances 
leads to Eq. \rf{sigma}.

\end{document}